\documentclass[twocolumn,superscriptaddress]{revtex4}
\usepackage{graphicx}
\usepackage{times}
\usepackage{amsmath}
\usepackage{amssymb}
\usepackage{units}
\usepackage{stmaryrd} 
\usepackage[compact]{titlesec}
\titlespacing{\section}{0pt}{*0}{*0}
\titlespacing{\subsection}{0pt}{*0}{*0}
\titlespacing{\subsubsection}{0pt}{*0}{*0}

\begin{document}
\preprint{0}

\title{Dirac states with knobs on:\\ interplay of external parameters and the surface electronic properties of 3D topological insulators}

\author{E. Frantzeskakis}
\email{e.frantzeskakis@uva.nl} 
\address{Van der Waals-Zeeman Institute, Institute of Physics (IoP), University of Amsterdam, Science Park 904, 1098 XH, Amsterdam, the Netherlands}

\author{N. de Jong}
\address{Van der Waals-Zeeman Institute, Institute of Physics (IoP), University of Amsterdam, Science Park 904, 1098 XH, Amsterdam, the Netherlands}

\author{B. Zwartsenberg}
\address{Van der Waals-Zeeman Institute, Institute of Physics (IoP), University of Amsterdam, Science Park 904, 1098 XH, Amsterdam, the Netherlands}

\author{T. V. Bay}
\address{Van der Waals-Zeeman Institute, Institute of Physics (IoP), University of Amsterdam, Science Park 904, 1098 XH, Amsterdam, the Netherlands}

\author{Y. K. Huang}
\address{Van der Waals-Zeeman Institute, Institute of Physics (IoP), University of Amsterdam, Science Park 904, 1098 XH, Amsterdam, the Netherlands}

\author{S. V. Ramankutty}
\address{Van der Waals-Zeeman Institute, Institute of Physics (IoP), University of Amsterdam, Science Park 904, 1098 XH, Amsterdam, the Netherlands}

\author{A. Tytarenko}
\address{Van der Waals-Zeeman Institute, Institute of Physics (IoP), University of Amsterdam, Science Park 904, 1098 XH, Amsterdam, the Netherlands}

\author{D. Wu}
\address{Van der Waals-Zeeman Institute, Institute of Physics (IoP), University of Amsterdam, Science Park 904, 1098 XH, Amsterdam, the Netherlands}

\author{Y. Pan}
\address{Van der Waals-Zeeman Institute, Institute of Physics (IoP), University of Amsterdam, Science Park 904, 1098 XH, Amsterdam, the Netherlands}

\author{S. Hollanders}
\address{Van der Waals-Zeeman Institute, Institute of Physics (IoP), University of Amsterdam, Science Park 904, 1098 XH, Amsterdam, the Netherlands}

\author{M. Radovic}
\address{Swiss Light Source, Paul Scherrer Institut, CH-5232 Villigen, Switzerland}
\address{SwissFEL, Paul Scherrer Institut, CH-5232 Villigen, Switzerland}

\author{N. C. Plumb}
\address{Swiss Light Source, Paul Scherrer Institut, CH-5232 Villigen, Switzerland}

\author{N. Xu}
\address{Swiss Light Source, Paul Scherrer Institut, CH-5232 Villigen, Switzerland}

\author{M. Shi}
\address{Swiss Light Source, Paul Scherrer Institut, CH-5232 Villigen, Switzerland}

\author{C. Lupulescu}
\address{Technische Universit\"{a}t Berlin, Institut f\"{u}r Optik und Atomare Physik, Strasse des 17. Juni 136, D-10623 Berlin,Germany}

\author{T. Arion}
\address{Institut f\"{u}r Experimentalphysik, Universit¬at Hamburg, Luruper Chaussee 149, 22761 Hamburg, Germany}
\address{Center for Free-Electron Laser Science / DESY, Notkestrasse 85, 22607 Hamburg, Germany}

\author{R. Ovsyannikov}
\address{Helmholtz-Zentrum Berlin f\"{u}r Materialien und Energie, Elektronenspeicherring BESSY II, Albert-Einstein-Strasse 15, 12489 Berlin, Germany}

\author{A. Varykhalov}
\address{Helmholtz-Zentrum Berlin f\"{u}r Materialien und Energie, Elektronenspeicherring BESSY II, Albert-Einstein-Strasse 15, 12489 Berlin, Germany}

\author{W. Eberhardt}
\address{Technische Universit\"{a}t Berlin, Institut f\"{u}r Optik und Atomare Physik, Strasse des 17. Juni 136, D-10623 Berlin,Germany}
\address{Center for Free-Electron Laser Science / DESY, Notkestrasse 85, 22607 Hamburg, Germany}

\author{A. de Visser}
\address{Van der Waals-Zeeman Institute, Institute of Physics (IoP), University of Amsterdam, Science Park 904, 1098 XH, Amsterdam, the Netherlands}

\author{E. van Heumen}
\address{Van der Waals-Zeeman Institute, Institute of Physics (IoP), University of Amsterdam, Science Park 904, 1098 XH, Amsterdam, the Netherlands}

\author{M. S. Golden}
\email{m.s.golden@uva.nl}
\address{Van der Waals-Zeeman Institute, Institute of Physics (IoP), University of Amsterdam, Science Park 904, 1098 XH, Amsterdam, the Netherlands}


\begin{abstract}
Topological insulators are a novel materials platform with high applications potential in fields ranging from spintronics to quantum computation.
In the ongoing scientific effort to demonstrate controlled manipulation of their electronic structure by external means - i.e the provision of knobs with which to tune properties - stoichiometric variation and surface decoration are two effective approaches that have been followed.
In angle resolved photoelectron spectroscopy (ARPES) experiments, both approaches are seen to lead to electronic band structure changes. Most importantly, such approaches result in variations of the energy position of bulk and surface-related features and the creation of two-dimensional electron gases.
The data presented here demonstrate that a third manipulation handle is accessible by utilizing the amount of super-band-gap light a topological insulator surface has been exposed to under typical ARPES experimental conditions. 
Our results show that this new, third, knob acts on an equal footing with stoichiometry and surface decoration as a modifier of the electronic band structure, and that it is in continuous and direct competition with the latter.
The data clearly point towards surface photovoltage and photo-induced desorption as the physical phenomena behind modifications of the electronic band structure under exposure to high-flux photons. We show that the interplay of these phenomena can minimize and even eliminate the adsorbate-related surface band bending on typical binary, ternary and quaternary Bi-based topological insulators.
Including the influence of the sample temperature, these data set up a detailed framework for the external control of the electronic band structure in topological insulator compounds in an ARPES setting. Four external knobs are available: bulk stoichiometry, surface decoration, temperature and photon exposure. These knobs can be used in conjunction to fine-tune the band energies near the surface and consequently influence the topological properties of the relevant electronic states.   
\end{abstract}

\maketitle

\section{INTRODUCTION}

Topological insulators (TIs) are a realization of a fascinating and novel state of quantum matter \cite{Hasan2010, Fu2007, Moore2010}.
Their remarkable properties result from metallic, spin-polarized surface states which possess linear dispersion: the so-called topological
surface states (TSS) \cite{Xia2009, Chen2009}.
Synergy between a chiral spin texture due to spin-momentum locking and time reversal symmetry \cite{Hsieh2009, Wang2010} means that the TSS are robust to non-magnetic impurities and disorder. Consequently, these systems are generating interest both in the context of spin generator systems in emerging spintronic devices \cite{Pesin2012} and as a platform for the creation of stable building blocks of quantum information in the form of Majorana zero modes \cite{Stern2013, Mourik2012}.
In order to realize such exotic applications, it would be beneficial to be able to selectively manipulate the electronic properties of the TSS. 
As viable alternatives to electrostatic gating (e.g. \cite{Chen2010_2, Zhang2013, Checkelsky2011, Kim2012}), extensive research has been carried out into two different means of effective manipulation, namely changes in the bulk stoichiometry of the TI material (e.g. \cite{Jia2011, Ren2010, Arakane2012, Neupane2012}) and controlled surface decoration of the TI crystal (e.g. \cite{Bahramy2012, King2011, Bianchi2011}). 

The requirement for high bulk resistivity in three dimensional TIs has led to a transition in terms of materials from binary teradymites (i.e. Bi$_{2}$Se$_{3}$, Bi$_{2}$Te$_{3}$ \cite{Xia2009, Hsieh2009, Chen2009})
to ternary systems
(i.e. [Bi$_{1-\textmd{x}}$Sb$_{\textmd{x}}$]$_{2}$Te$_{3}$ \cite{Zhang2011, Kong2011}),
and finally, to TlBiSe$_{2}$-based \cite{Kuroda2010, Sato2010, Chen2010} and quaternary Bi$_{2}$Te$_{2}$Se-based \cite{Ren2010, Jia2011, Lin2011, Neupane2012} compounds (i.e. Tl$_{1-\textmd{x}}$Bi$_{1+\textmd{x}}$Se$_{2-\textmd{y}}$ \cite{Kuroda2013} and Bi$_{2-\textmd{x}}$Sb$_{\textmd{x}}$Te$_{3-\textmd{y}}$Se$_{\textmd{y}}$ \cite{Ren2011, Arakane2012, Pan2014}, respectively).
In these systems, $x$ and $y$ serve as the parameters to fine tune the energy position of the TSS with respect to the bulk electronic states and the Fermi energy.
On the other hand, the surface engineering route to effective tuning of the electronic band structure has been achieved by decorating the surface of a TI crystal or film with metals (via deposition) \cite{Zhu2011, Valla2012, King2011, Wray2011, Bahramy2012}, via controlled gas adsorption \cite{Bianchi2011, King2011, Benia2011, Chen2012PNAS, Jiang2012}, or by means of exposure to the residual gases in a UHV environment (i.e. time-dependence in vacuum) \cite{Bianchi2010, Zhu2011, Benia2011,King2011}. 

To enable TIs to reach their full potential, the complete parameter space for band structure control and manipulation should be investigated and understood.
By contrast to the plethora of studies following the two scenarios sketched above, only very few reports have been made regarding variations of the electronic band structure based on other external parameters such as temperature variations or exposure to continuous illumination \cite{Jiang2012,Kordyuk2011}.

This paper reports a detailed study of the interplay between three {\it in-situ} factors affecting the surface band structure, namely illumination, surface decoration and temperature. The analysis considers variations of the bulk conductivity over orders of magnitude via comparison of binary, ternary and quaternary 3D TIs.

Using Angle-Resolved PhotoElectron Spectroscopy (ARPES) at different experimental setups and under variable experimental conditions (see Methods), we demonstrate that the amount of illumination the surface region has received is not a mere secondary factor in determining the surface band structure of Bi-based TIs.
On the contrary, it will be evident that exposure to photons can completely cancel the desired modifications induced by deliberate stoichiometric tuning of the bulk and/or induced by controlled surface decoration.
Besides describing the observed photo-induced effects and discussing the underlying mechanisms, the investigations reported here provide the necessary context in which past results on TIs acquired by techniques involving photon exposure -such as ARPES- need to be viewed. In doing so, these results also provide a framework for future research.

In the following, we will first present the effects of residual gas adsorption and exposure to illumination with a variable flux for the case of Bi$_{2}$Se$_{3}$.
We then present analogous UHV adsorption and photon flux experiments on two, related, quaternary TIs with high bulk resistivity: Bi$_{1.5}$Sb$_{0.5}$Te$_{1.7}$Se$_{1.3}$, Bi$_{1.46}$Sb$_{0.54}$Te$_{1.7}$Se$_{1.3}$.
The third data section deals with the stoichiometric compounds Bi$_{2}$Te$_{2}$Se and BiSbTeSe$_{2}$.
The main body of the paper finishes with a discussion of the underlying physical phenomena and a summary of the main consequences of our work for past and future band structure studies of these materials.
The paper additionally contains two appendices in which the effect of sample temperature is discussed, for both pristine cleavage surfaces exposed to UHV and for intentionally metal-decorated surfaces.\\

\section{RESULTS: The effect of illumination}

\subsection*{Bi$_{2}$Se$_{3}$}

Fig. 1 illustrates our findings on Bi$_{2}$Se$_{3}$, the most widely studied topological insulator.
In the top panel, we present the energy position of (i) the spin degeneracy point of the TSS (also known as the Dirac point) and (ii) the onset of the valence band leading edge at $k=-0.11\textmd{\AA}^{-1}$, both plotted as a function of time.
In comparison to past studies \cite{King2011, Bianchi2010, Zhu2011}, we investigate the effect of an additional parameter which turns out to be of great impact: the incident photon flux.
For times up to the point represented in Fig. 1a by the solid vertical line at \textit{ca.} 500 minutes, the photon flux was kept at a `low' value: $F_{\textmd{low}}$$= 1.3$$\times$10$^{19}$ photons$/(\textmd{s}$ $\textmd{m}^{2})$ (see Methods).
In this regime, both the Dirac point and the valence bands can be seen to shift by the same amount to higher binding energies as time progresses, signalling a rigid shift of the electronic band structure. 
This phenomenon has been widely reported as due to downward band bending (BB), as a result of a near-surface electrostatic potential \cite{Bahramy2012, Bianchi2010, King2011, Zhu2011, Bianchi2011}.
The width of the space charge region that accompanies the downward BB has been found to be as large as 20-30 nm from ARPES \cite{Bianchi2011, King2011, Benia2011} and HAXPES studies \cite{ViolBarbosa2013}.
As the data in the right-hand panel of Fig. 1a show, the time evolution of an increasing downward BB can be reversed when - all other parameters being held constant - the photon flux in increased by a factor of 100.
Upon such strong illumination $[$flux $F_{\textmd{high}}$$=$100$\times$$F_{\textmd{low}}$$= $1.3$\times$10$^{21}$ photons$/(\textmd{s}$ $\textmd{m}^{2})$$]$, both the TSS and near-surface bulk-derived bands shift back to lower binding energies, and do so initially at a significantly faster rate than the initial, downward shift.
From the above, it is clear that the complete picture in an ARPES experiment is highly-dependent on the total photon fluence to which the sample is exposed; the result is a dynamic competition between the downward band bending in UHV and a photo-induced flattening of the bent bands.

\begin{figure}[!b]
  \centering
  \includegraphics[width = 8.5 cm]{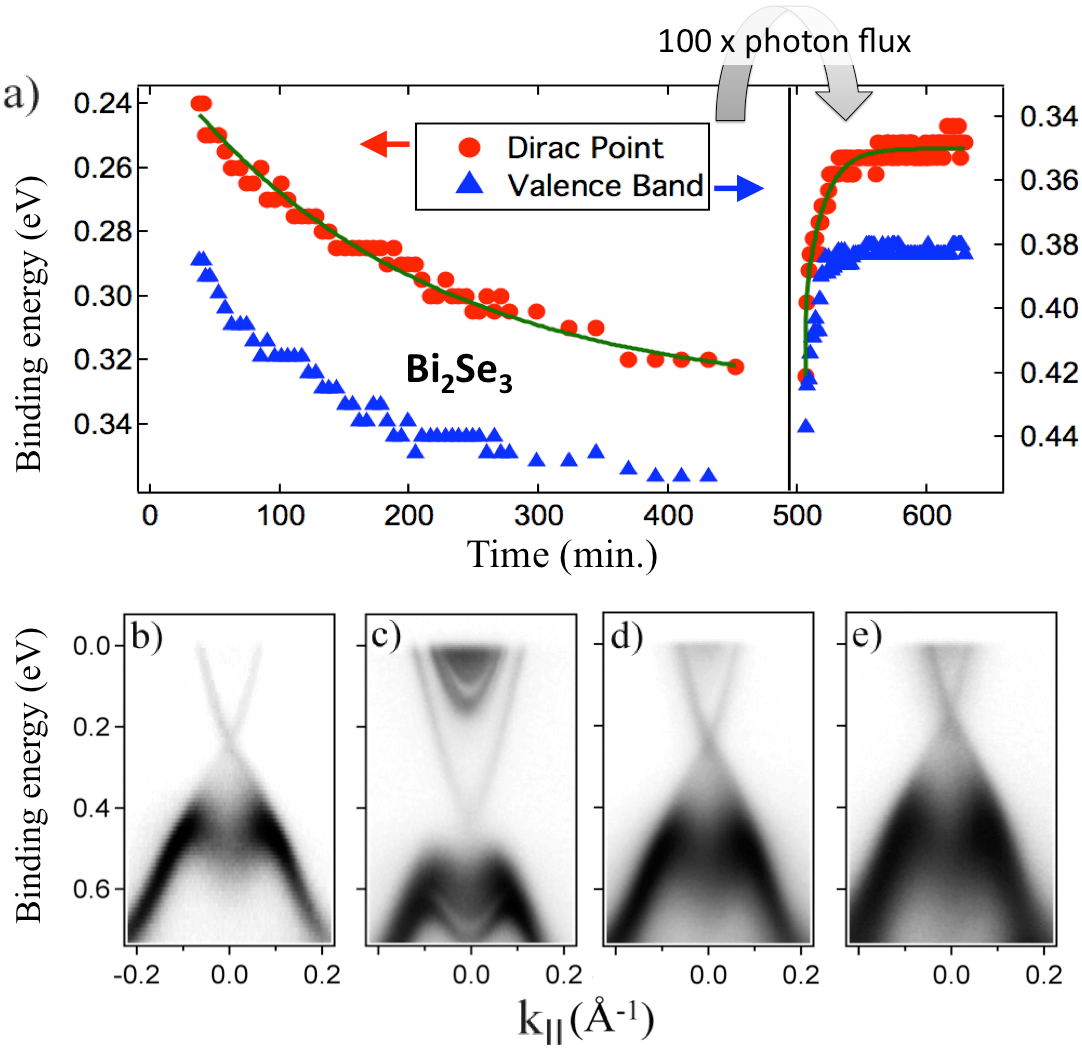}
  \caption{\textbf{
Time-dependent changes in the electronic band structure of Bi$_{2}$Se$_{3}$ under weak and strong illumination}
(a) Time-dependent energy position of the Dirac point (left axis) and the leading edge of the VB (right axis).
The photon flux is increased by a factor of 100 at the time marked by the vertical line.
Time intervals refer to the time since cleavage of the crystal in UHV.
The error bars in the value of binding energy are estimated to be $\pm5$ meV for both the Dirac point and VB.
The green solid lines represent single (left panel) and double (right panel) exponential functions used to determine the time constants (see Eqn. 1) for the downward band bending and the fast(slow) upward reversal thereof under high-flux illumination.   
(b)-(d) The near-E$_{\textmd{F}}$ electronic band structure of Bi$_{2}$Se$_{3}$ at different time intervals and beam exposures: (b) a freshly cleaved sample without any prior exposure to the photon beam [i.e. the first point on curve (a)], (c) I($E,k$) image acquired after 11 hours from an area of the same sample that received only a limited amount of illumination, (d) same location on the sample as (b) but now after exposure to both the low and high-flux beams [i.e. last point on the curve in the right-hand panel of (a)], (e) same location on the sample as (c) after additional exposure to higher energy photons ($h\nu=130$ eV) until saturation is observed.
All data have been acquired at 16K. 
}
\label{fig1}
\end{figure}

According to the most widely accepted scenario, the built-in electrostatic potential at the origin of the downward BB in binary TIs is driven by the decoration of the sample surface with residual gas atoms which act as electron donors on chemisorption \cite{Benia2011, King2011, Bianchi2011, Bianchi2010}.
Strongly-developed BB gives rise to a potential well in which also the bulk conduction and valence band (CB and VB) become quantized \cite{Bahramy2012, King2011, Bianchi2011}.
Quantum well states in the valence band have been alternatively suggested to arise from the expansion of the van der Waals spacing between consecutive quintuple layers of the TI through the intercalation of gases
\cite{Eremeev2012, Ye2011, Bianchi2012}.
In either case, the effect of residual gas atoms is central to time-dependent band structure modifications in UHV, which are sometimes referred to as aging in the ARPES literature.

An alternative scenario for the downward BB in Bi$_{2}$Se$_{3}$ invokes the formation of Se vacancies leading to the formation of Se atoms in the gas phase and two extra electrons per vacancy in the crystal \cite{Cava2013}.
In such a case, one would expect that the bands of Se-terminated Bi$_{2}$Se$_{3}$ would shift faster than Bi$_{2}$Te$_{2}$Se, which is predicted to be Te-terminated \cite{Cava2013}.
We did not observe a difference in behaviour between these two systems in our experiments.
Moreover, as discussed later, the downward shift is more pronounced at low sample temperatures, which would be somewhat counterintuitive in the Se-vacancy scenario.
For the above reasons, we also attribute the downward BB to the effect of residual gas atoms.

We observe no or at most marginal differences between the time-dependent energy shifts of the surface- and bulk-related states (Fig. 1a).
This is a natural consequence of the high surface-sensitivity of our ARPES data.
Using $h\nu=23$ eV, we probe the part of the CB and VB localized within 1 nm of the surface termination \cite{Seah1979}; a value which is smaller than the out-of-plane spatial confinement of the bulk bands in the band-bending potential, and smaller than the spatial extension of the surface state in the $z$-direction, which is 2-3 nm \cite{Zhang2010}.
  
We attribute the dramatic changes on the sign and rate of the observed energy shifts on high-flux illumination to the combined effect of two physical phenomena:
\begin{itemize}
\item surface photovoltage (SPV) effects
\item photo-induced desorption of adsorbed species
\end{itemize}
SPV effects are commonly encountered in semiconductor materials upon illumination with photons of energy exceeding the bulk band gap.
Light absorption results in the formation of electron-hole pairs via interband transitions.
These are separated in the space charge region via the built-in electric field that is responsible for the downward band bending. The spatial distribution of the SPV-induced charge carriers in the band-bending potential opposes the original charge distribution that gave rise to the band bending. In this way, these photo-induced charges cancel the original electrostatic potential and thus lead to a decrease in the band bending \cite{Kronik1999, Alonso1989, Demuth1986, Lagowski1973, Long2002,Leibovitch1994}. Insight into the delicate synergy of SPV effects and photo-stimulated
desorption of foreign species from the sample surface will be given in Section III. Here we note that surface- and bulk-related features shift rigidly to lower binding energy $(E_{\textmd{b}})$ and that the size of the photo-induced energy shift is a function of the total photon fluence.
The bottom line is thus a continuous competition between the adsorbate-related charge redistribution and the photo-induced effects; i.e. between the effects of the residual vacuum (downward BB) and of the high-flux illumination (upward shift in energy, neutralizing the same downward BB). 

The realization that there is a continuous competition between adsorbate-related and photo-induced changes clearly has consequences for how ARPES can be used to extract information on the energy position of the surface and bulk-related features in Bi-based 3D TIs.
Any such determination is incomplete without explicit reference to
\begin{enumerate}
\item the residual gas pressure
\item the time-window the cleaved surface has been in that specific vacuum
\item the photon fluence (i.e. total number of photons received per unit area).
\end{enumerate}

In particular, the last of these three factors has barely been addressed in the Bi$_{2}$Se$_{3}$ ARPES literature: illumination effects have been considered in a single study that focused on $h\nu$-dependent changes \cite{Kordyuk2011}. 

The energy ($E$) \textit{vs.} wavevector ($k$) dispersion for a fresh surface of Bi$_{2}$Se$_{3}$ (recorded within 35 min. after cleavage) is shown in the I($E,k$) image of Fig. 1b.
The Dirac point lies at a binding energy of 240 meV - the first red data point in Fig. 1a - and the bulk conduction band rates minimum is located just below E$_F$, although not readily visible for the greyscale used (see also Fig. 2d).
For such `low' flux as the one used for the left-hand panel of Fig. 1a, the effect of residual gas adsorption prevails over the potential flattening and/or desorption induced by illumination and thus the binding energy of the Dirac point increases to 320 meV after 400 min.
As can be seen in Fig. 1c, at sample positions of the same cleave which have not been continuously exposed to the photon beam, the binding energy of the Dirac point has even higher values. 
At such locations, the BB is now strong enough to act as a potential well and clear signatures of confinement are then observed in the conduction and valence bands \cite{King2011, Bahramy2012, Valla2012}. 
The observation that unexposed areas are different reveals a degree of local character in the photon-induced changes on Bi$_2$Se$_3$, which is dramatically enhanced for bulk-insulating quaternary TI compounds \cite{Frantzeskakis2014}.
 
The data shown in Fig. 1d correspond to the last data point of Fig. 1a ($t=630$ min).
Under higher flux illumination, the Dirac point has now shifted back to a binding energy equivalent to that in panel (b). 
Occupation of the CB states with electrons is signalled by the spectral weight close to the Fermi level visible inside the linearly dispersing TSS.
As the I($E,k$) image of Fig. 1e shows, the band structure can be shifted to even lower binding energies via additional exposure to higher energy photons ($h\nu$=130 instead of 23 eV).
Comparing panels (d) and (e) with panel (b), one may observe that
the apparent energy difference between the CB minimum and the Dirac point decreases with increasing $h\nu$ exposure, while the bands themselves appear broadened. These two observations are intimately connected as the measured image is a superposition of different areas under the beam-spot, each of which is up-shifted by a differing amount.

The `high' photon flux used to record the (up-shifting) data shown in the right-hand panel of Fig. 1a is of the same order as the standard flux that a sample receives at modern synchrotron facilities operating in multi-bunch mode with a typical storage ring current of 400 mA.
The data presented in Fig. 1a were acquired under such conditions at the Swiss Light Source, and the same effects have been reproduced at the BESSY II source at the Helmholtz Zentrum Berlin.
For us to observe the well-studied, downward BB and the evolution of quantum well states in Bi$_{2}$Se$_{3}$ \cite{King2011, Bahramy2012, Zhu2011, Benia2011, Bianchi2010}, the photon flux had to be deliberately decreased by a factor of 100, as was done for the data in the left-hand panel of Fig. 1a.
With the benefit of hindsight, therefore, it is no surprise that the pioneering studies on the time-dependent, downward BB of TIs were performed with low flux sources, namely laboratory setups \cite{Zhu2011, Benia2011} $[$photon flux $\sim$10$^{18}$ photons$/(\textmd{s}$ $\textmd{m}^{2})$$]$ \cite{Greber1997, Souma2007}, low-flux synchrotron sources \cite{Bianchi2010, Bianchi2011} or large-scale facilities operated at low storage ring current such as is the case in so-called single-bunch mode \cite{King2011}.
Under standard, high flux conditions (400 mA ring current; insertion device beamline), the illumination effects are so strong at low temperature that they can essentially prevent the time-dependent, downward BB.\\

\subsection*{Quaternary compounds with high bulk resistivity}

In bulk crystals, the transport properties of the topological surface states in Bi$_{2}$Se$_{3}$ and Bi$_{2}$Te$_{3}$ are difficult to investigate, as bulk conduction due to intrinsic impurities and crystallographic defects dominates. 
Although progress has been achieved on charge carrier doping \cite{Analytis2010,Hor2009}, thin film engineering and electrostatic gating \cite{Chen2010_2,Checkelsky2011}, an alternative has been explored in the development of bulk materials with reduced bulk conduction. The quaternary system Bi$_{2-\textmd{x}}$Sb$_{\textmd{x}}$Te$_{3-\textmd{y}}$Se$_{\textmd{y}}$ (BSTS) compound \cite{Arakane2012} has been at the forefront of these efforts.
We have also investigated the effect of low- and high-flux illumination on crystals of two BSTS compositions: Bi$_{1.5}$Sb$_{0.5}$Te$_{1.7}$Se$_{1.3}$ (BSTS1.5) and   (BSTS1.46).
In contrast to Bi$_{2}$Se$_{3}$, both BSTS1.5 and BSTS1.46 exhibit insulator-like resistivity curves with typical low temperature values of order 10 Ohm$\cdot$cm (Fig. 2a). The low-tempearature resistivity of these quaternary compounds is more than 4 orders of magnitude higher than the resistivity of Bi$_{2}$Se$_{3}$.\\
\begin{figure}
  \centering
  \includegraphics[width = 8.7 cm]{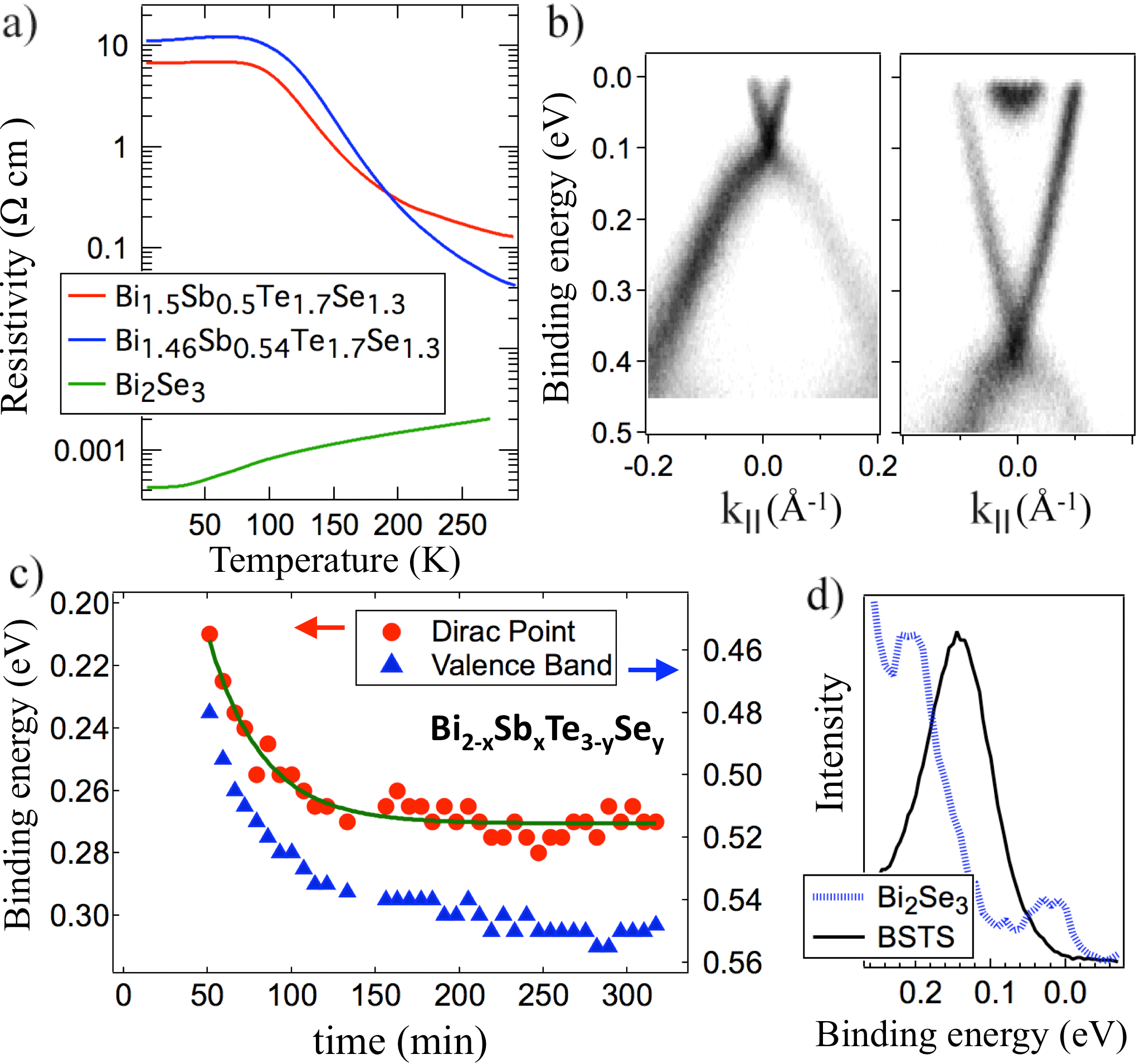}
  \caption{\textbf{The effect of low-flux illumination on quaternary TIs with high bulk resistivity: Bi$_{1.5}$Sb$_{0.5}$Te$_{1.7}$Se$_{1.3}$ (BSTS1.5) and Bi$_{1.46}$Sb$_{0.54}$Te$_{1.7}$Se$_{1.3}$ (BSTS1.46)} (a) $T$-dependent resistivity of BSTS1.5,  BSTS1.46 and Bi$_{2}$Se$_{3}$. (b) The near-E$_{\textmd{F}}$ electronic band structure of BSTS1.46 after only minimal exposure to photons and UHV exposure as indicated. (c) BSTS1.5: Time-dependent energy position of the Dirac point (left axis) and the intensity maximum of the $k$-integrated valence band spectrum within the energy window of measurement (right axis) under weak illumination $[$1.3$\times$10$^{19}$ $/(\textmd{s}$ $\textmd{m}^{2})$$]$. Error bars in the value of binding energy are estimated to be $\pm10$ meV for the Dirac point and $\pm5$ meV for the valence band.
The green solid line represents an exponential fit to the time-dependence of the downward band bending of the Dirac point (Eqn.1).   
(d) Comparison of the energy distribution curves at $k_{\|}=0$ for freshly cleaved BSTS1.5 and Bi$_{2}$Se$_{3}$ (both 35-40 min. after cleavage), revealing the absence of a low binding energy peak signalling the conduction band in BSTS. The Bi$_{2}$Se$_{3}$ spectrum is from the data shown in Fig. 1b. All ARPES data were acquired at 16K from single crystals from the same batch as those used for the resistivity measurements.
}
\label{fig2}
\end{figure}

\subsubsection*{Low flux regime}

The binding energy of the Dirac point for freshly cleaved BSTS1.46 and BSTS1.5 samples typically ranges between 0.1 and 0.2 eV (Fig. 2). This uncertainty arises from the fact that the downward BB
sets in at its maximum rate in the first minutes after cleavage (Fig. 2c).
Unlike the case of Bi$_2$Se$_3$, cleaved BSTS1.5 and BSTS1.46 samples are characterized by the complete absence of a bulk conduction band at the Fermi level, as seen in the near-$E_{\textmd{F}}$ energy dispersion (Fig. 2b, after 0.6 h in UHV). The comparison shown in Fig. 2d between the $k_{\|}=0$ energy distribution curves for freshly cleaved Bi$_{2}$Se$_{3}$ and BSTS1.5 (the former data trace is a vertical slice through the data of Fig. 1b), showing that electron occupy the conduction band states only in the binary TI.

However, downward band bending in BSTS can also bring the bulk conduction band well below $E_{\textmd{F}}$, as is seen in the right-hand panel of Fig. 2b for BSTS1.46.
By tracking the complete time dependence (Fig. 2c) of the surface band structure of BSTS1.5 at low photon flux (flux $F_{\textmd{low}}$), the first signs of CB occupation are seen within a time-scale of 100 min for these vacuum conditions (i.e. $P=5\times10^{-11}$ mbar) at low temperature ($T=16$ K).
The data shown in Figs. 2b and 2c are therefore a vivid demonstration that also in BSTS-based compounds, time-dependent BB should also be taken into account when discussing the energy position of the Dirac point with respect to E$_{\textmd{F}}$.
The data shown in Figs. 1 and 2 illustrate that variation in the time interval between cleavage and measurement can yield changes in the observed binding energies that are greater than the intentional changes made via alteration of the chemical composition of the bulk crystals.

A comparison of the early data points in Figs. 1a and 2b reveals that the downward band bending evolves faster in BSTS than in Bi$_{2}$Se$_{3}$ under equivalent conditions.
This means that while BSTS reaches a saturated energy shift after approximately 150 mins of UHV exposure to a low photon flux, the analogous, downward energy shift in Bi$_{2}$Se$_{3}$ fails to saturate within the time window of our measurement.
The saturation energy value for the BB in BSTS shown in Fig. 2c represents the outcome of a continuous competition between decoration with residual gas atoms and exposure to the low-flux photon beam. Just as in the case of Bi$_{2}$Se$_{3}$ (Fig. 1c), at sample locations that have been only minimally exposed to photons, greater band bending is observed.

In order to provide a quantification of the energy shifts, we have fitted an exponential function to the time-dependent data shown in Figs. 1a and 2c:
\begin{equation}
E(t) = \pm E_{0} e^{-t/\tau}
\end{equation}
with a positive prefactor describing the downward BB (in which the Dirac point binding energy increases with time) and a negative prefactor, the upward shift under high flux conditions.
The characteristic time constant $\tau$ for the band bending is 193 min for Bi$_{2}$Se$_{3}$ and 32 min for BSTS, the latter echoing the `steeper' temporal evolution of the downward shift in BSTS. 
A possible origin for this behaviour could lie in the more effective screening of the adsorbate-induced surface charges underlying the downward BB in Bi$_{2}$Se$_{3}$, compared to the bulk-insulating BSTS.
We also note that as the chemical composition of the termination layer of the two compound differs, this could also impact the sticking coefficient for residual gas atoms, and therefore also modify the BB behavior.\\

\subsubsection*{High flux regime}

Fig. 3 shows the influence of a high-flux photon beam on BSTS. High flux means $F_{\textmd{high}}$$= 1.3$$\times$10$^{21}$ photons$/(\textmd{s}$ $\textmd{m}^{2})$. The starting point is a saturated downward BB surface. Under this strong illumination, the binding energy of the Dirac point is seen to reduce, indicating that the bands flatten and the overall band structure is shifting upwards to lower binding energy (Fig. 3a).
As was observed for Bi$_{2}$Se$_{3}$ (right panel of Fig. 1a), this photon-induced effect is rather rapid in the first few minutes of high flux exposure.

The I($E,k$) images in panels (b), (c) and (d) of Fig. 3 show the observed band structure changes in BSTS on increasing high-flux beam exposure, in which pronounced spectral broadening is apparent, taking the form of a shadow-like feature surrounding the main experimental bands. As in Bi$_{2}$Se$_{3}$, the broadening is a manifestation of the local action of the photon beam in which distinct electronic band structures are summed, each up-shifted differently in energy due to the different number of photons in the spatially inhomogeneous profile of the photon beam. 
The data of Fig. 3 show the local character of these photon-induced effects to be further enhanced in the case of bulk-insulating BSTS.
This effect has been modelled in Ref. \onlinecite{Frantzeskakis2014}, and exploited to create micro-metric patterns in the topological band structure.

\begin{figure}
  \centering
  \includegraphics[width = 8.6 cm]{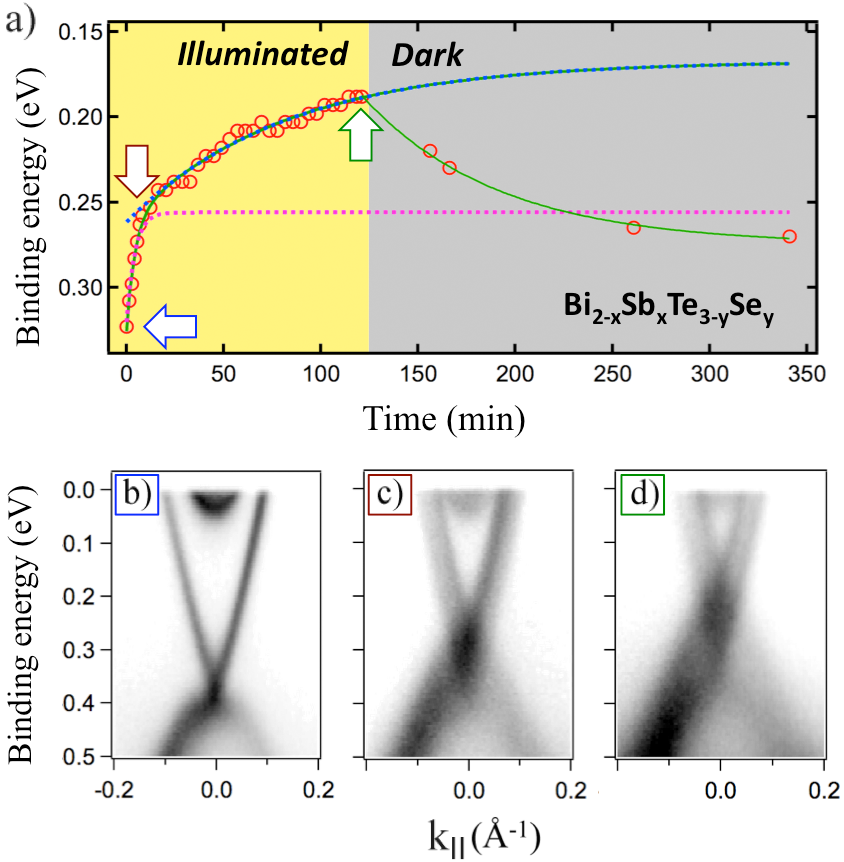}
  \caption{\textbf{The effect of high-flux illumination on quaternary TIs with high bulk resistivity: Bi$_{1.5}$Sb$_{0.5}$Te$_{1.7}$Se$_{1.3}$ (BSTS1.5) and Bi$_{1.46}$Sb$_{0.54}$Te$_{1.7}$Se$_{1.3}$ (BSTS1.46)} (a) BSTS1.46: Time-dependent energy position of the Dirac point (open circles) under the influence of strong illumination (yellow panel) and after continuous illumination ceased (grey panel). The energy position of the Dirac point recovers only partially once the light is off, and does so on a much longer time-scale than the onset of the photo-induced upward shift in the yellow panel. 
Green solid lines are exponential fits on the data points using Eqn. 1. As described in the text, the `illuminated'  region is best fitted with a double exponential function whose constituent curves are shown by pink and blue dashed lines. A single exponential fit can accurately describe the `dark' region.
Arrows in panel (a) [i.e. BSTS 1.46] denote the time intervals where the analogous data on BSTS1.5  [panels (b) - (d)] has been acquired. The error bars in the value of binding energy are between $\pm5$ and $\pm15$ meV. (b) BSTS1.5: near-E$_{\textmd{F}}$ electronic dispersion upon minimal photon exposure after considerable UHV exposure (i.e. 10 hours after cleavage). (c) Same as (b) but after 15 min high-flux photon exposure $[$1.3$\times$10$^{21}$ photons$/(\textmd{s}$ $\textmd{m}^{2})$$]$. (d) Same as (c) but with the total high-flux exposure time now increased to 2 hours. All data have been acquired at 16K.
}
\label{fig3}
\end{figure} 

\begin{figure} [!b]
  \centering
  \includegraphics[width = 8.67 cm]{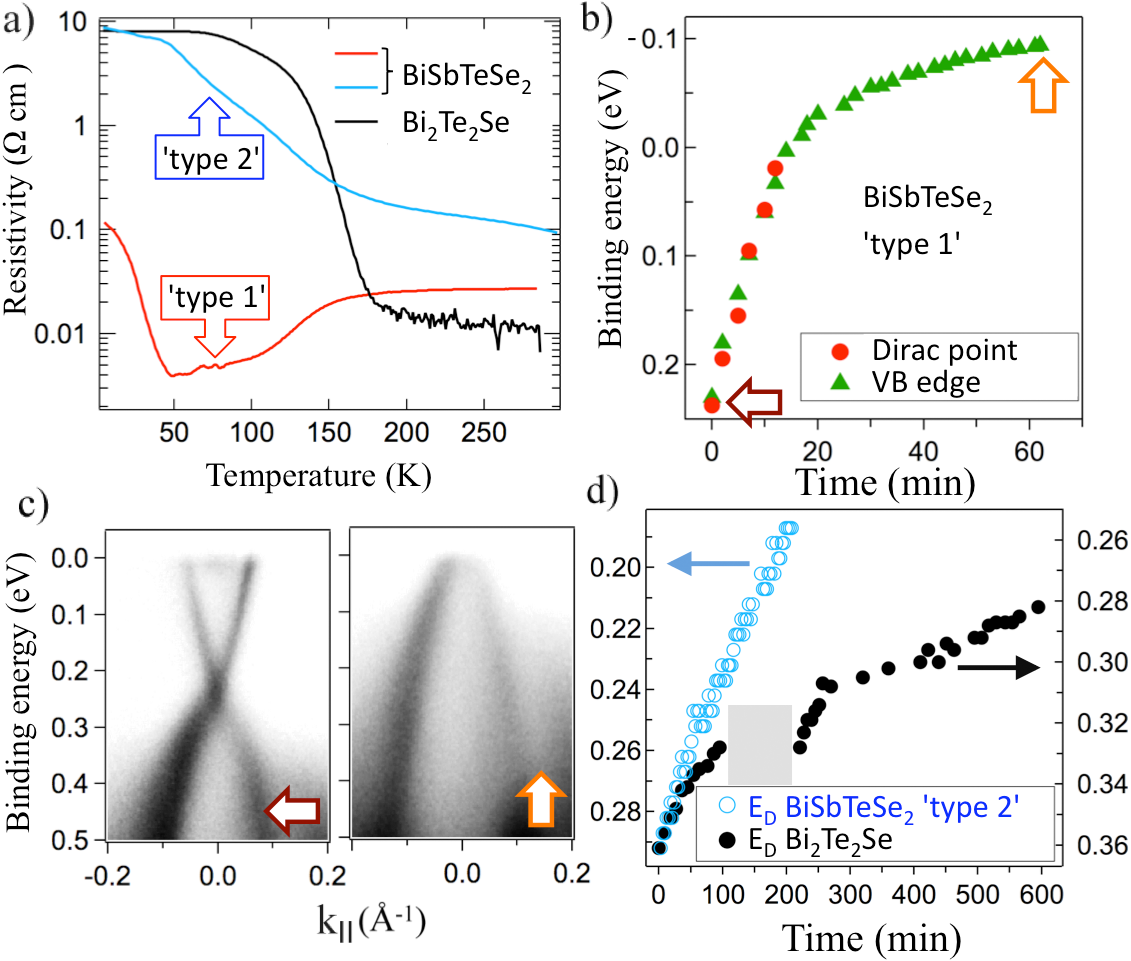}
  \caption{\textbf{The effect of high-flux illumination on Bi$_{2}$Te$_{2}$Se (BTS) and BiSbTeSe$_{2}$ (BSTS2)} (a) $T$-dependence of the resistivity of BSTS2 (both `type 1' and `type 2' $[$for definitions, see text$]$) and BTS. (b) Time-dependent energy position of the Dirac point of type 1 BSTS2 under the influence of a high-flux photon beam, with data from the valence band edge superimposed as described in the main text  (c) BSTS2: near-$E_{\textmd{F}}$ electronic band structure for a type 1 sample at times as indicated by the arrows in panel (b). (d) Time-dependent energy position of the Dirac points of BSTS2 (`type 2', blue symbols) and BTS (black symbols) under the influence of a high-flux photon beam. The grey area denotes a time-period in the BTS experiment that was without high-flux photon exposure. All data have been acquired at 16K. Resistivity and ARPES measurements were performed on the same crystal pieces.}
\label{fig4}
\end{figure}

Unlike the downward BB in Figs. 1a (left panel) and 2c, the time evolution of the {\it upward} energy shift on high-flux illumination cannot be accurately fitted with a single exponential.
For both Bi$_{2}$Se$_{3}$ and BSTS a double exponential function is required.
The resulting fits are shown using solid green lines in the left-hand panel of Fig. 3a for BSTS and in the right-hand panel of Fig. 1a for Bi$_{2}$Se$_{3}$.
For both materials, the double exponential fits contain a `fast' time constant ($\tau_1$) that describes the first few minutes of the upward shift, and a `slow' time constant ($\tau_2$) that tracks the evolution at later stages.
For the BSTS data shown in Fig. 3, the $\tau_1$($\tau_2$) curves are shown in Fig. 3a using pink(blue) dashed lines.
In Section III, we bring together all the exponent fit-data in Table 1, and discuss how the $\tau$ values in the double-exponents are consistent with the combined effects of surface photovoltage (faster) and slower photo-induced desorption processes.

The final step in the presentation of the ARPES data from BSTS is the temporal decay of the upward SPV-induced energy shift - i.e. the return to a time-depedent {\it increase} in the binding energies of the Dirac point and other characteristic features of the electronic structure - when the high-flux illumination source is turned off.
The right panel of Fig. 3a shows that on high-flux switch-off, the downward BB begins to reassert itself, and the topological surface state returns to higher binding energy, a process which can be captured by a third exponential relation shown in the figure by the solid, green line.  

In analogy to the situation in conventional semiconductors, we attribute the sluggish nature of this return to downward BB to the very small drift velocity across the space charge region (SCR), meaning that photo-induced charges can exist well beyond the exposure time, depending on temperature and on the magnitude of the original band bending. Lifetimes of photo-induced charges of up to $10^{7}$ seconds have been reported for GaAs \cite{Hecht1991}.\\

\subsection*{Other ternary and quaternary compounds}

Before Section III presents a discussion of the relative merits of SPV and photon-stimulated desorption as mechanisms behind the reversal of the downward BB under high-flux illumination, the data collection feeding this discussion will be completed with data from other ternary and quaternary 3D TI compounds.

Fig. 4 summarizes the data for the two nominally stoichiometric compounds Bi$_{2}$Te$_{2}$Se (BTS) and BiSbTeSe$_{2}$ (BSTS2).
Panel (a) shows the temperature-dependence of the resistivity for these compounds.
BTS shows a similar resistivity curve to the two BSTS compositions studied in Figs. 2 and 3, and the low-T saturation value of the resistivity is only 20\% smaller than for BSTS1.46.

Resistivity curves of BSTS2 were found to vary significantly from sample to sample.
More than 20 samples have been measured and two characteristic cases which were widely encountered are shown in Fig. 4a. 
BSTS2 samples of what we refer to here as `type 1' (red trace in Fig. 4a) retain an insulating up-turn in resistivity, but the low-T value is 2 orders of magnitude lower than for the BSTS samples (Fig. 2a) and BTS (Fig. 4a, black trace). Moreover, for `type 1' BSTS2 samples the resistivity exhibits a clear transition to a more metallic behavior at higher temperatures.
The overall, temperature-dependent resistivity curve of `type 1' BSTS2 is reminiscent of that in extrinsic semiconductors in which, as the temperature is raised, the resistivity initially decreases due to the thermal excitation of electrons trapped at impurity sites followed by a high-temperature increase as the carrier mobility decreases due to thermal fluctuations.
In short, we attribute differences in the resistivity curve of BSTS2 samples of `type 1' and other ternary and quaternary compounds to the larger bulk defect density of the former. On the other hand, BSTS2 samples we refer to as `type 2' (blue trace in Fig. 4a) show an insulator-like T-dependence of the resistivity at all temperatures.
The low-T resistivity value of `type 2' BSTS2 samples is very similar to BTS (Fig. 4a) and other BSTS samples (Fig. 2a). The ARPES signatures of the two different types of BSTS2 samples after exposure to a high-flux photon beam are very instructive for establishing a link between the bulk carrier concentration and their response to external illumination in this system. 

Figs. 4b-4d present our ARPES results on BTS and BSTS2 after exposure to illumination comparable to the high-flux beam used in Figures 1 and 3.
The electronic structure of `type 1' BSTS2 samples is very sensitive to external illumination. Fig. 4b shows that the Dirac point was shifted upward to the Fermi level after only 17 minutes of high-flux exposure.
To be able to determine the trend at later times, as a first step, an estimate of the density of states is made by integrating the I($E, k$) images along the momentum direction.
The shift of the valence band edge in these effective DOS data is then tracked as a function of time and the resulting binding energy vs. time curve is re-scaled, so it matches that from the Dirac point over their common range.
The data of panel (b) show a turnover to a slower evolution shortly after the Dirac point crosses $E_F$. Although the data do not show a saturation of the upward shift, it is clear from the I($E, k$) images of `type 1' BSTS2 shown in Fig. 4c that for the highest exposures measured, the Fermi level is definitely below the Dirac point, pointing to the $p$-type character of this material.
Fig. 4d shows that the BSTS2 samples with large low-T resistivity (`type 2') also show an upshift of the Dirac point upon high flux illumination, but this takes place slower than for the `type 1' crystals, and in the time window for which there are data, there are no signs of a slowing or saturation of the up-shift.

The overall behavior of BTS on increasing photon fluence (Fig. 4d black markers) is similar to `type 1' BSTS2 shown in panel (b), but the magnitude and rate of the upward shift are much smaller in BTS. The shaded area in panel (d) denotes a time period for which the photon beam illuminated other (neighboring) locations on the BTS sample.
During this time we estimate that only a small amount of radiation affected the sample location from which these ARPES data points have been obtained.
Interestingly, the rate of the {\it initial} upward shift in BTS - i.e. that which would be governed by a time-constant $\tau_1$ - is very similar to that of `type 2' BSTS2 (Fig. 4d empty blue symbols).
As none of the binding energy vs/ time curves in Fig. 4 saturate, we refrain from carrying out (double-)exponential fits of the temporal behavior.

Summarizing the main points arising from the data presented so far, it is clear that competing trends in the energy position of the topological surface states, as well as the valence and conduction bands in Bi-based 3D topological insulators result from UHV exposure (linked to adventitious adatom adsorption) and high-flux photon beam exposure.
This implies that the energy position of the near-surface electronic band structure of different TI compounds can be fine-tuned, after the bulk stoichiometry of the bulk crystal has been determined in the growth process.
In the next section, a discussion of the possible mechanisms behind the observed changes is presented, followed by examination of insight that can be gained by considering the sample temperature as a further parameter.\\

\section{DISCUSSION}

\subsection*{Two control `knobs'}

In the above, we have identified and characterized two deterministic, controllable and (given long timescales) reversible parameters (or `knobs'), each of them operative after the synthesis of the topological insulator material in question.
The first control parameter is the well-known, adsorbate-induced band bending \cite{Benia2011, Cava2013, Bianchi2011}, which leads to an increase of the Dirac point binding energy.
An alternative interpretation of these observations could be band bending due to the intrinsic properties of the surface of what is an extrinsic (doped) semiconductor.
However, for $n$-type Bi$_{2}$Se$_{3}$ \cite{Cava2013, WangAPL}, this mechanism would predict upward BB \cite{Alonso1989, Kronik1999}, the reverse of what is observed experimentally \cite{Bahramy2012, King2011}.
Therefore the assignment to adsorbate-induced BB is the more convincing of the two.

A second external control parameter is based on the interaction of the near-surface region of the topological insulators with a photon beam.
This interaction leads to a decrease of the Dirac point binding energy, the energy value of which can be tuned via choice of the photon flux, energy and exposure time. This second knob operates at a local spatial level \cite{Frantzeskakis2014}.\\

\subsection*{The physical phenomena induced by high-flux illumination}

The data suggest that the underlying mechanism behind the response to high-flux illumination is a combination of the surface photovoltage (SPV) effect and of the photo-induced desorption of adsorbed atoms.
Before discussing the delicate interplay between these factors in more detail, we discuss some alternative mechanisms.
\begin{itemize}
\item \textit{Negligible local heating from the high-flux beam}. Calculation of the  local, beam-induced temperature rise of the sample for a 23 eV photon beam with an output of 1.3$\times$10$^{21}$ photons$/(\textmd{s}$ $\textmd{m}^{2})$ and with a Gaussian profile of $\textmd{FWHM}=30$ $\mu \textmd{m}$ (typical experimental conditions), arrives at maximally a few mK \cite{Lax1977}. This effectively excludes beam-induced heating as the origin of the observed upward shifts of the bands.
\item \textit{Photo-induced creation of vacancies is not the dominant process}. We observe the same kind of photo-induced, upward shifts on Se-terminated Bi$_{2}$Se$_{3}$, as for Te-terminated Bi$_{2}$Te$_{2}$Se, and for systems with mixed Se-Te termination such as BSTS \cite{Cava2013}. This observation cannot easily be consistent with the creation of Se (or Te) vacancies on the surface of TIs after exposing them to high-flux photon beams.
\end{itemize}

Apart from the above mentioned mismatches for alternative scenarios, the observation of two different time-scales for the upwards energy shifts (fitted using double exponentials) would seem to fit to an interplay of SPV and photo-induced desorption phenomena. 
SPV is an electronic effect and can be reasonably expected to be faster than the desorption of adatoms.
Since the $\tau_1$ values extracted (see Tab. 1) describe processes some twenty times faster than $\tau_2$, it would be natural to associate $\tau_1$ with SPV and $\tau_2$ with photo-induced desorption.
However, it should be stressed that a {\it combination} of SPV and desorption is necessary to describe the reported data of Figs. 1-4. In particular:
\begin{itemize}
\item The `few-hour' timescale observed in Figs. 3 and 4 for the saturation of the photo-induced upward shifts goes well beyond anything reported for surface photo-voltage effects in regular semiconductors such as Si or GaAs. In these systems, SPV has been shown to saturate on time-scales ranging between only picoseconds to microseconds \cite{Hecht1991,Long1990}, so the long-timescle effects seen for these TI surfaces can also involve photon-stimulated desorption. 
Having said that, we do note that SPV time-scales of a few minutes, comparable to the $\tau_1$'s found in this study, have been reported for the SPV onset in ZnO nanorods \cite{Zhao2007}.
\item The sensitivity of the photo-induced upward shift to the photon energy (see Fig. 1d \textit{vs.} 1e: greater for higher photon energy) does argue for a role for the SPV-mechanism and is in keeping with a recent report on Bi$_{2}$Se$_{3}$ \cite{Kordyuk2011}.
\item The temperature dependence of the energy shifts presented in Appendix 1 of this paper show qualitative agreement with the expectations for a process in which SPV plays a significant role.
\end{itemize}

One aspect in which the observed behavior differs from the standard SPV expectation regards the Fermi level. 
SPV-induced charges accumulating at the surface of a semiconductor should lead to the deviation of the Fermi level from the equilibrium position \cite{Kronik1999}, yet we do not observe this. This suggests that also for ternary and quaternary Bi-bassed 3D TIs the topological surface states are important in maintaining charge neutrality at the surface, as they have been argued to be in Bi$_{2}$Se$_{3}$ \cite{Kordyuk2011}.
A difference between the bulk-insulating systems and Bi$_{2}$Se$_{3}$ (with its degenerate, defect-doped bulk) is that it is possible that photo-induced charges can be trapped locally in the sub-surface region. When swept apart by the in-built potential which gives rise to the downward band bending, these charges may therefore alter the energetics of the surface electronic structure in a manner somewhat analogous to a local back-gate, without the need for explicit surface charging and the requirement of a measurable shift of the Fermi level at the surface from the equilibrium position \cite{Kordyuk2011}.\\

\subsection*{The challenge of obtaining flat band conditions}

The data of Figs. 1 and 2 show that low-flux ARPES spectra acquired on binary, ternary and quaternary Bi-based TIs correspond to a band-bent situation
\cite{Bianchi2010, Zhu2011, Benia2011, Bianchi2011, King2011},
as the initial, downward shift evolves very quickly in the first few minutes after cleavage.
It is therefore questionable whether flat band conditions can be captured experimentally within this short time window after cleavage.
Although the data of Figs. 1-4 suggest that photo-induced effects could be used to neutralize the downward band bending, before any conclusions are drawn as regards the true, flat-band electronic structure of the TI systems studied, it is appropriate to discuss briefly the differences between considering an extrapolation of low-flux or high-flux data to the limit of infinite fluence:
\begin{itemize}
\item {\bf Low flux case.} If the exponential fits made to model the time dependence of the Dirac point energy - for example in the left-hand panel of Fig. 1a or in Fig. 2c - were used to extrapolate to `infinite fluence', it is clear that the $F_{\textmd{low}}$ case cannot yield flat band conditions. This underlines that there is more afoot than only the creation of (very) long-lived electron hole pairs in the reversal of the downward band bending. 
\item {\bf High flux.} The time dependences observed in Figs. 1a (right), Fig. 3a (left) and Fig. 4b for Bi$_2$Se$_3$, BSTS and BSTS2 show that the extrapolation to `infinite fluence' using $F_{\textmd{high}}$ conditions yields a principally different end-point than the  $F_{\textmd{low}}$ case. The high-flux limit is clearly much closer to flat band conditions than the low-flux case.   
\end{itemize}

\begin{figure} 
  \centering
  \includegraphics[width = 8.7 cm]{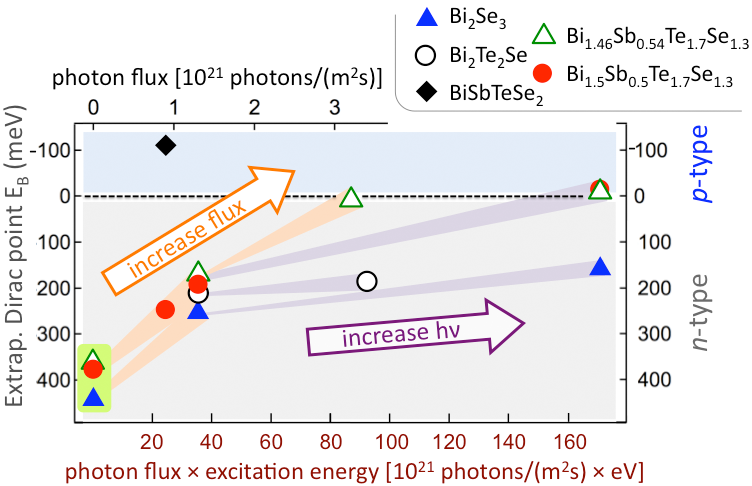}
  \caption{
  \textbf{The effect of photon flux and excitation energy in approaching flat band conditions} The vertical axis is the minimum binding energy (E$_B$) of the Dirac point, arrived at by infinite extrapolation of exponential fits to the time dependence on high-flux illumination. Symbols represent the E$_B$ values for different 3D TI compounds. The left-most data points underlaid in green show the maximally band-bent starting situation before high-flux illumination. The top axis gives the photon fluxes; and the bottom axis the product of photon flux and photon energy.
The orange guides to the eye show how the Dirac point binding energy decreases as the flux is increased and the purple guides to the eye illustrate further band flattening when, already at high flux, the photon energy (and hence the product flux$\times$$h\nu$) is increased. 
The right-most data-point of both BSTS samples and the BSTS2 sample lie on the light blue background, and thus, from a close-to-flat-band ARPES perspective can be considered to be \textit{p}-type. On the other hand, for Bi$_2$Se$_3$ and BTS the analogous Dirac point E$_B$ remains on the grey background, making these materials \textit{n}-type.}
\label{fig5}
\end{figure}

\begin{center}
\begin{table*}[ht]
\caption{\textbf{From band bent (BB) to flat band conditions - compilation of key data for various 3D topological insulators.}
This table brings together ARPES data on the electronic structure of a number of Bi-based TI systems after downward BB, and on how flat-band conditions for these systems can be generated using high-flux photon exposure. These data are compared to the low-T electrical resistivity of each system.
{\bf Row 1:} the initial binding energy of the Dirac point ($E_{\textmd{D}}$) is extracted from the figure panels indicated which track the evolution of the upward energy shift on high-flux illumination.
{\bf Row 2:}  time constants for the double-exponential fit to the high-flux-induced {\it upward} shift of the electronic structure to lower binding energy.
{\bf Row 3:} flat-band $E_{\textmd{D}}$ values. Minimum binding energy of the Dirac point, from extrapolation of the exponential fits to the time dependence when the product flux$\times$$h\nu$ has its highest value for each compound (i.e. right-most data points for each symbol in Fig. 5). 
{\bf Row 4:} this 5 minute time window for up-shifting corresponds to that described by $\tau_1$, where double-exponential fits have been made.
Shown is the upshift as a percentage of the difference between the values shown in Rows 1 and 3. 
High degrees of BB-reversal are encountered at short exposure times in samples showing relatively low resistivity. {\bf Row 5} $\rho$ at T=16K.
{\bf Row 6:} for the designation {\it n}$|${\it p}-type, the Fermi level - under flat band conditions - is located at the edge of the bulk CB$|$VB, as derived from the data of Row 3 and the I($E,k$) images from ARPES.   
{\bf Row 7:} largest observed $E_{\textmd{D}}$ values, presumably under saturated (downward) band bent conditions (see also Fig. 6). 
{\bf Row 8:} gives the maximum value of downward BB observed on exposure to residual UHV gases, given by the difference between the data of Row 7 and the flat-band $E_{\textmd{D}}$ values of Row 3.}
    \begin{tabular}
    {c @{\hskip 0.2cm} c @{\hskip 0.4cm} c @{\hskip 0.4cm} c @{\hskip 0.4cm} c @{\hskip 0.3cm} c @{\hskip 0.4cm} c @{\hskip 0.2cm}}\\
        \hline \hline
        \textbf{No.} &\textbf{Quantity} &\textbf{Bi$_{2}$Se$_{3}$} &\textbf{BiSbTeSe$_{2}$} &\textbf{Bi$_{1.46}$Sb$_{0.54}$Te$_{1.7}$Se$_{1.3}$} &\textbf{Bi$_{2}$Te$_{2}$Se} & \textbf{BiSbTeSe$_{2}$}\\ 
     & & & \textbf{(type 1)} &  &  & \textbf{(type2)}\\ \hline \hline
    1. & Initial $E_{\textmd{D}}$ (meV binding energy) & 325 & 235 & 325 & 360 & 290\\
    & This paper, figure panel & 1a & 4b & 3a & 4d & 4d\\ \hline
   2. & $\tau_1$, $\tau_2$ (both in min.), on exposure to $F_{\textmd{high}}$ & 0.6, 14.8 &  & 4.3, 81.8 &  & \\ \hline
     3. & $E_{\textmd{D}}$ (meV binding energy) at flat bands & 160 & -110 & -10 & 185 & -110\\ \hline \hline
   4. & \% of initial, downward BB reversed after 5 min. at  $F_{\textmd{high}}$ & \:\:\:69\% & \:\:\:52\% & \:\:\:13\% & \:\:\:1.4\% & \:\:\:0.6\% \\ \hline
   5. & Electrical resistivity (T=16K, in Ohm$\cdot$cm) & $<$0.001 & $<$0.1 & 11 & 8 & 8 \\ \hline \hline
   6. & \textit{n}- or \textit{p}-type at flat band conditions? & \textit{n}-type & \textit{p}-type & \textit{p}-type & \textit{n}-type & \textit{p}-type\\ \hline
   7. & Max. $E_{\textmd{D}}$ (meV binding energy), {\it via} UHV exposure & 440 & 290 & 360 & 470 & 290 \\ \hline
   8. & Max. BB (meV, downward), {\it via} UHV exposure & \:280 & \:400 & \:370 & \:285 & \:400\\ \hline \hline
                      \end{tabular}
     \end{table*}
 \end{center}
 
Taking this point to heart, Fig. 5 illustrates how the best achievable approximation of flat band conditions can be generated using external, high flux illumination.
The vertical axis shows the binding energy of the Dirac point obtained by extrapolation of the double exponential functions that describe the upward shift of the bands (see for example Fig. 3a $[$yellow background$]$) to infinite time.
The starting point is the left-most group of symbols (superimposed on a green background). These are not extrapolated values, but are the Dirac point energies from  I($E,k$) images showing the most downwardly band bent situations encountered, prior to high-flux illumination.  
The first step towards flat-band conditions is through increasing the photon flux as shown in the top horizontal scale.
This leads to an evolution of the extrapolated Dirac point binding energies as shown by the orange guides,.

For the BSTS2 sample (solid black diamond in Fig. 5), this is already enough to put the extrapolated Dirac point more than 100 meV above the Fermi level, clearly indicating this material to be - in the absence of downward band-bending - \textit{p}-type.
For the other materials, with maintenance of the high flux, the photon {\it energy} was increased.
The bottom axis shows the product of the flux and the photon energy, and this number can be seen to represent the ability of the high-flux photon beam to create (multiple) electron-hole pairs, freeing charges to flatten the band bending at the surface of the crystal.
The extrapolated Dirac point energies of the BSTS samples (green triangle, red filled circle) then shift further to still lower binding energy (see purple guides), finally landing just above the Fermi level for the highest value of the product of flux and photon energy.
This suggests that like the case of BSTS2, the surface region of the BSTS-based materials would be \textit{p-type}, in the absence of downward band bending.
In contrast, the extrapolated Dirac point binding energies of Bi$_2$Se$_3$ (blue triangle) and BTS (white circle), remain steadfastly more than 150 meV below the Fermi level, indicating these systems are - also without downward bend bending - \textit{n}-type. 

The combination of high photon flux and higher photon energy that yields the right-most symbols for the Dirac point energy for each TI compound in Fig. 5 represents as close to flat band conditions as can be practically achieved in an ARPES experiment.
Table 1 gathers together the key electronic structure data from the different materials investigated in this manner.
Even though the strategy adopted seems successful in removing the distorting effects of the near-surface downward band-bending, it should still be recognized that the real-life situation is still a dynamic equilibrium between the upward-shifting effects of the photon flux (SPV, photon-stimulated desorption) and the downward-shifting adsorbate-induced band-bending.\\

\subsection*{Comparison of the response of the various TI compounds to high-flux illumination}
In Section II, it was described how high-flux illumination can reverse adsorbate-related downward BB and in Section III the pro's and con's of the different physical phenomena were reviewed in favor of a combination of surface photo voltage and desorption effects.
Table 1 brings together not only the key electronic structure data for the maximally bend bent condition but also those relevant to the `flat band' situation.
In addition, Tab. 1 shows the rate at which the band bent situation was reversed, expressing the up-shift within the first five minutes of high-flux illumination in terms of a percentage of the original (downward) band bending. 
In particular, it is clear that the crystals with low resistivity at the cryogenic temperatures at which the ARPES data in the main body of the paper were recorded  - and that means Bi$_2$Se$_3$ and the `type 1' crystals of BSTS2 -  exhibit a fast upshift of between 50 and 70 \% of the original band bending.
In particular, comparison of the two types of BSTS2 crystal is interesting, whereby the fast upshift of the `type 2' samples (which have a high resistance at low T) is less than 1\% of the original downward band bending.

\begin{figure*}
  \centering
  \includegraphics[width = 17.7 cm]{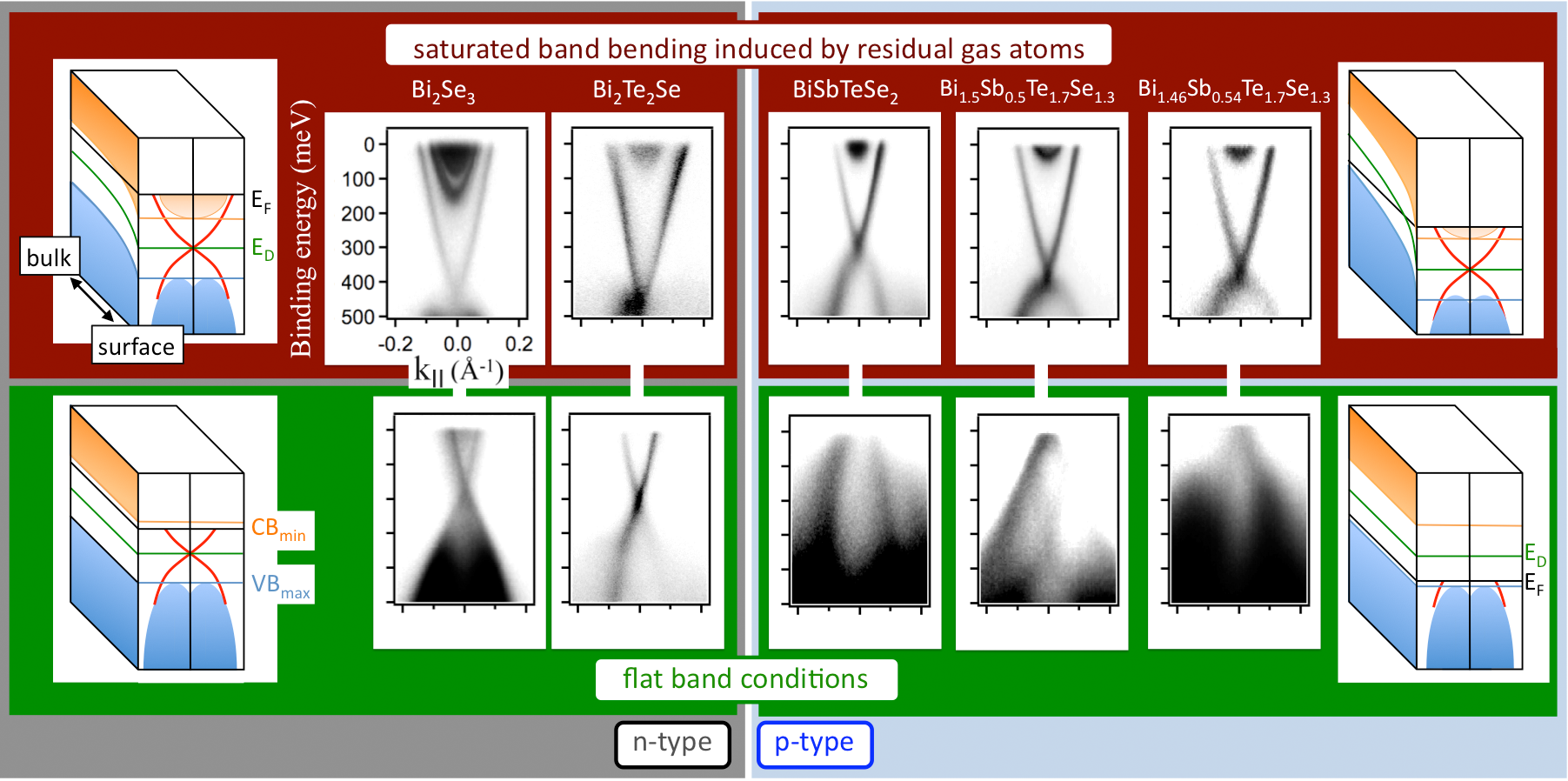}
  \caption{\textbf{From saturated UHV-induced band bending to flat band conditions by high-flux illumination.} (Top row) Experimental I($E,k$) images arrived at by exposure of different compounds to residual gases in UHV ($P<2\times10^{-10}$mbar) for time periods in the range of 10-48 hours, during which external illumination of these specific sample locations was kept to a minimum and the downward band bending was seen to saturate (see also row 7 of Table 1).
(Bottom row) Experimental I($E,k$) images recorded after high-flux, high photon-energy exposure, ensuring close to flat-band conditions. The energy of the Dirac point for each data panel in the lower row differs by less than 20 meV from last (right-most) data point for each corresponding material plotted in Fig. 5 (see also row 3 of Table 1). Samples whose data are on a grey (light blue) background correspond to \textit{n}-type (\textit{p}-type) compounds and their Fermi energies are pinned within a few meV of the bottom of the conduction band (top of the valence band). The four schematic illustrations show the electronic band structure for band bending (top row) and flat-band (bottom row) conditions, for both \textit{n}-type (left) and \textit{p}-type (right) compounds.
Data have been acquired at temperatures between 16K and 35K, and all I($E,k$) images are plotted on the same axes and scales as Bi$_2$Se$_3$.}
  \label{fig6}
\end{figure*}

Assuming a simple Schottky model for which the maximal BB exceeds the thermal broadening, the solution of Poisson's equation provides a relationship between the band bending potential $(V)$ and the width of the space charge region $(d)$:
\begin{equation}
V=\frac{e N_{b} d^{2}}{\varepsilon \varepsilon_{0}}
\end{equation}
where $N_{b}$ is the bulk carrier density, $\varepsilon$ is the static dielectric constant and $\varepsilon_{0}$ is the vacuum permittivity \cite{luth-book-2010}. 

If the bulk carrier density increases, as is the case on going from the `type 2' to the `type 1' BSTS2 crystals, one would expect the the width of the space charge region (SCR) to decrease for a given total band bending potential. 
By means of parametric solutions of all equations related to the band bending and charge transfer, recently Brahlek \textit{et al.} provided contour plots showing the relation between the SCR and $N_{bulk}$ for 3D topological insulators \cite{Brahlek2014}.

If the SCR for the `type 2' BSTS2 and the other highly resistive Bi-based TIs were indeed broader, this could slow the carrier diffusion across the SCR, and in turn this would slow the charge redistribution leading to the flattening of the downward BB - regardless of whether SPV or stimulated desorption of adsorbates is at work. 

Within this framework, the width of the space charge region in BSTS1.46 ($N_{b}$$=$$2\times10^{15}$ cm$^{-3}$) can be expected to be more than 10 times larger than for Bi$_{2}$Se$_{3}$ ($N_{b}$$=$$5\times10^{17}$ cm$^{-3}$), taking the bulk carrier concentrations from Ref. \onlinecite{Pan2014}. 
Although this model is too crude to capture the details of the adsorbate-induced band bending, it provides important insight which is also in line with the results presented here.
The transport data tell us that Bi$_{2}$Se$_{3}$ has the most bulk carriers, followed by `type 1' BiSbTeSe$_{2}$.
Correspondingly, these are the two Bi-based TI compounds we studied that exhibit the fastest rates for the reversal of the band bending (see Figs. 1a and 4b), which would be in keeping with their possession of a narrower SCR.
In contrast, the more resistive Bi$_{1.46}$Sb$_{0.54}$Te$_{1.7}$Se$_{1.3}$,  Bi$_{1.5}$Sb$_{0.5}$Te$_{1.7}$Se$_{1.3}$, `type 2' BiSbTeSe$_{2}$  and Bi$_{2}$Te$_{2}$Se have a broader space charge region and display more sluggish BB reversal rates on high photon flux exposure.

An independent indication of the relative extension of the SCR is given in Fig. 1c and the right panel of Fig. 2b. Clear confinement effects are seen in the electronic band structure of Bi$_{2}$Se$_{3}$ after 11 h UHV exposure (Fig. 1c), while no signs of quasi-2D confinement is seen at all in the measured electronic structure of BSTS1.46, even after exposure to the same UHV conditions for 48 hours (Fig. 2b, right). Again, this speaks for a narrower SCR in  Bi$_{2}$Se$_{3}$, compared to BSTS.

The energy position of the Fermi level deep inside the bulk of a 3D topological insulator is determined by its bulk carrier density, $N_{b}$ \cite{Brahlek2014}. When $N_{b}$ is smaller than $10^{18}$ cm$^{-3}$, $E_{\textmd{F}}$ is pinned close to the conduction band minimum for an \textit{n}-type material or similarly close to the valence band maximum for a \textit{p}-type system \cite{Brahlek2014}. For the surface region of either type of material - and it should be stressed that it is this region of the sample that is relevant for ARPES, STM/STS or transport experiments for true bulk insulating samples - the same holds {\it only} if flat band conditions can be generated.  
Consequently, given that the high-flux, high photon-energy ARPES data are as close as one can come to flat band conditions in the context of a surface sensitive experiment (Table 1, row 3), the spectroscopic data presented here can be used to sort the Bi-based, 3D TI compounds we have studied into either \textit{n}-type or \textit{p}-type electronic structures in the near-surface region. This is done in the data shown in Table 1.

To bring home this important point in a graphical manner, Figure 6 shows - for each of five different Bi-based 3D topological insulators studied - the juxtaposition of the I($E,k$) image for the maximally downward BB `starting' situation, with the analogous I($E,k$) image for the most up-shifted situation observed, recorded using high-flux illumination.
The top row of Fig. 6 is also the state of affairs one would expect to be confronted with in the moderate vacuum encountered in cryogenic apparatus used to measure transport properties of bulk crystals.  
Under such conditions, the conduction bands of all of the Bi-based 3D TIs we have investigated are partially occupied - they possess \textit{n-type} surfaces, with topologically trivial states mixed in with the TSS.
The lower row of Fig. 6 essentially shows the situation for flat bands at the surface.
These I($E,k$) images show the most `upshifted' band structures we have observed. These experimentally measured data differ by at most 20 meV from the values resulting from the time-infinite extrapolation shown as the right-most symbols for each material type in Fig. 5.

Comparison of the upper and lower row of I($E,k$) images in Fig. 6 brings the differing alignments of the bands and the Fermi level sharply into focus.
For the $n$-type compounds (Bi$_2$Se$_3$ and Bi$_2$Te$_2$Se), the bands move such that the Fermi level goes from being well into the conduction bands (BB, see the top-left sketch of the electronic structure), to only just grazing them (flat-band, see the bottom-left sketch) \cite{footnote}.
For the $p$-type compounds (BSTS2, BSTS1.5, BSTS1.46), the bands change such that the Dirac point moves from in excess of 250 meV binding energy (with the conduction bands clearly cutting the Fermi level, as shown in the top-right sketch of the band alignment), to E$_D$ lying above $E_F$, such that the valence band maximum is close to the Fermi level in the flat-band situation (sketched in the lower-right schematic).

Consequently, Fig. 6 is a nice illustration that once an effective flat-band-like situation can be reached, the Fermi level is either close to the bottom of the bulk CB or just above the top of the bulk valence band, as would be expected from straightforward semiconductor physics considerations.\\

\section{CONCLUDING REMARKS}

All in all, our results offer new insight into an ongoing topic of great current interest, namely the optimization of the energy position of the topological surface states with respect to the Fermi level and with respect to the bulk bands in three dimensional topological insulators.
Effective optimization strategies can be viewed as the first step towards future, application-oriented research.
The new results, combined with the advantage of hindsight, enable us to arrive at an interpretative framework into which both past experimental work should be placed and the context in which future studies can be viewed:

{\bf (a) Fast downward band bending in UHV.}
From an ARPES point of view, the tunability of the energy position of the TSS cannot be solely accomplished by the bulk stoichiometry.
3D topological insulator compounds with bulk stoichiometries chosen so as to give real bulk insulating transport characteristics (e.g. BSTS1.5 and BSTS1.46) are subject to adsorbate induced band bending at their surfaces just as their binary counterparts \cite{Bahramy2012, King2011, Bianchi2011,Zhu2011,Benia2011,Valla2012}.
Adsorbate induced band bending in quaternary compounds evolves quickly after cleavage and details of the band structure measured using ARPES depend as much if not more on the time interval between cleavage and measurement, than on the composition itself.
When the downward band bending saturates, a well-resolved conduction band is seen to cross the Fermi level at low temperature, thus although the bulk may be insulating, the surface region possesses more than only the TSS conduction channel.

{\bf (b) Strong effect of illumination.} 
The effect of a photon beam on the surface electronic structure of Bi$_{2}$Se$_{3}$, ternary and quaternary Bi-based TI compounds has been investigated and found to be considerable for fluxes typical at 3$^{\textmd{rd}}$ generation synchrotron radiation sources, given an adsorbate-induced downward band bent situation as the starting point before illumination.
The direct and significant effect of the photon beam underlines that great care needs to be taken when using ARPES to infer quantitative trends in the energy position of the various features in the electronic band structure.
Not only do the lifetime of the cleave and the background pressure (strictly also residual gas chemical composition) need to be logged, but also the photon flux and the irradiation time of the sample location under investigation.
These findings may enable the generation of quantitative agreement between previous studies \cite{Bianchi2010, Zhu2011, Benia2011,King2011}. Interestingly, the value of photon flux used to effectively cancel the adsorbate-induced, downward band bending is comparable to that on offer in regular laboratory laser and LED sources $[$1.3$\times$10$^{21}$ photons$/(\textmd{s}$ $\textmd{m}^{2})$$]$.

{\bf (c) Combination of surface photovoltage and desorption phenomena.}
The marked effects of high-flux VUV illumination on the energy landscape of band-bent Bi-based 3D TIs are attributed to the combined effects of surface photovoltage and photon-stimulated desorption phenomena. The combination of an electronic process (SPV) and a mass-transport-related process (desorption) fits the dual `fast' and `slow' (twenty times slower than the `fast') characteristic timescales observed for the photon-beam-related upshift of the electronic states. In Appendix 1, raising/re-lowering the sample temperature is shown to reversibly decrease/increase the downward band bending across the whole sample, and is due to temperature-induced desorption/re-adsorption at the surface. Appendix 2 shows that on a qualitative level, the upward energy shift of key features of the electronic structure that accompanies high-flux induced band flattening does not depend on the microscopic origin of the band bending, and also that the efficacy of this band-flattening is reduced at elevated temperature. Both these observations on the data presented in Appendix 2 are in keeping with the importance of surface photovoltage in the observed phenomena.  

{\bf (d) High-flux / high-energy illumination yields flat-band conditions at the surface.}
As the onset of downward band bending on adatom adsorption (even in UHV) is very rapid, using a standard ARPES experiment to reliably determine the binding energies (relative to the Fermi level) of the Dirac point, as well as the valence and conduction bands in the near-surface region is not practicable.
In contrast, extrapolation of the energies determined in high-flux, ARPES experiments conducted at higher photon energies is able to yield reasonable values for the relative alignment of the key features of the electronic structure, due to the effective flattening of the bands at and near the surface under such conditions. The higher is the flux $\times$ $h\nu$ product, the closer is the extrapolated energy value to flat-band conditions.

{\bf (e) Reliable ARPES determination of \textit{n}- or \textit{p}-type character at low temperatures.}
Use of extrapolated high-flux (and/or high-$h\nu$) binding energies (i.e. flat-band conditions) shows Bi$_2$Se$_3$ and Bi$_2$Te$_2$Se to be \textit{n}-type. On the other hand, BiSbTeSe$_2$, and the two high-resistivity BSTS compositions studied [Bi$_{1.5}$Sb$_{0.5}$Te$_{1.7}$Se$_{1.3}$ and Bi$_{1.46}$Sb$_{0.54}$Te$_{1.7}$Se$_{1.3}$] are \textit{p}-type materials.\\

Taken together, points (a)-(e) show how a full understanding of the energetics of the ARPES data from Bi-based 3D topological insulators can be arrived at, once explicit consideration is taken of the fact that illumination is an integral part of the experiment.
Not only can high-flux ARPES data be used for a reliable determination of the character of the majority carriers in the TI system, but they also point to the possibilities of illumination-based, in-situ tuning of the key features of the electronic structure.
Looking forwards, judicious use of super-band-gap illumination at a flux level readily achievable with laboratory light sources could be used as a valuable and reversible tool to decide whether additional topologically trivial surface-confined bulk states are at the chemical potential, contributing to transport, or whether future device-related experiments can be carried out using bulk crystals in which only the topological surface states count.\\    

\section*{METHODS}

\subsection*{Sample growth} Crystals were grown in Amsterdam using the Bridgman technique.
High purity elements (Bi [99.999~\%], Sb [99.9999~\%], Te [99.9999~\%] and Se [99.9995~\%]) were melted in evacuated, sealed quartz tubes at 850$^{\textmd{o}}$C and allowed to mix for 24 hours before cooling.
The cooling rate was 3$^{\textmd{o}}$C per hour. Samples were cleaved in UHV ($P<5\times10^{-10}$mbar) and at temperatures between 16-38K.\\

\subsection*{Angle resolved photoelectron spectroscopy} ARPES experiments were performed at four different experimental setups:

(i) At the SIS-HRPES end-station of the Swiss Light Source with a Scienta R4000 hemispherical electron analyzer. The minimum sample temperature was 16K and the pressure during measurements was $5\times10^{-11}$ mbar.

(ii) At the UE-125 NIM beamline of BESSY II using the IDEEAA end-station \cite{Lupulescu2013} with a VG Scienta ArTOF 10k electron analyzer. The minimum sample temperature was 20K and the pressure during measurements was better than $5\times10^{-10}$ mbar.

(iii) At the UE112-PGM-2a-1$^{2}$ end-station of BESSY II with a Scienta R8000 hemispherical electron analyzer. The minimum sample temperature was 35K and the pressure during measurements was $2\times10^{-10}$ mbar.

(iv) In Amsterdam using a Scienta VUV5000 helium photon source and a Scienta 2002 hemispherical electron analyzer. The minimum sample temperature was 19K and the pressure during measurement was $2\times10^{-10}$ mbar (this is a pressure of ultra-pure He gas, the base pressure without operating the light source was $3\times10^{-11}$ mbar).

Data relevant to the competition of adsorbate- and photo-induced shifts were measured with $h\nu=23$ eV and $h\nu=27$ eV, while the polarization of the photon beam was linear horizontal. Exposure to high-energy light was done with $h\nu=70$ eV and $h\nu=130$ eV. After exposure to high-energy photons the I($E,k$) images were recorded  with $h\nu=23$ eV or $h\nu=27$ eV. The photon flux was deliberately decreased from $F_{\textmd{high}}$ (SIS and UE112 sources only) to $F_{\textmd{low}}$ by detuning the undulator gap, while keeping the nominal monochromator energy and exit slit - and thus photon footprint at the sample - fixed.
Relative values of the photon flux were deduced by comparing with current values from a standard photodiode. 
The energy position of the Fermi level was determined within 5 meV by measurements of an evaporated Au thin film that was held in electrical contact with the sample. \\

\section*{ACKNOWLEDGEMENTS}
We thank Thomas Pichler for useful discussions.
This work is part of the research programme of the Foundation for Fundamental Research on Matter (FOM), which is
part of the Netherlands Organization for Scientific Research (NWO).
EvH acknowledges support from the NWO Veni program.
The research leading to these results has also received funding from the European CommunityÕs Seventh Framework Programme
(FP7$/$2007-2013) under grant agreement n$^{\textmd{o}}$ 312284 (CALIPSO).

\newpage

\section*{APPENDIX 1: the effect of sample temperature on BSTS1.5}

The data in this first part of the appendix show that changing the sample temperature provides a fully reversible approach by which the electronic properties at the surface can be changed on a global spatial level. 
Fig. A1a presents the energy position of the Dirac point and the CB minimum of a bulk-insulating BSTS1.5 crystal as a function of raising the sample temperature.
The photon flux used for the measurement is low enough that conventional, downward BB is observed when the sample is continuously irradiated in UHV at fixed $T$. 
Fig. A1b shows the time-dependence of the relevant band energies for a different BSTS1.5 sample while it thermalises after rapid heating of the cryostat from 20K to 150K.
It is clear from both panels that the effect of increasing sample temperature is to decrease the downward BB, as has been reported by others \cite{Zhu2011,Jiang2012}.

Fig. A1a shows that the shift of the bands to lower binding energy can be reversed again, simply by reducing the temperature once more. The open circles for T=20 K are the result of such a re-cool process. The small green arrows in Fig. 7a show that after re-cooling, the binding energies of both the Dirac point and the CB continue to increase with time. 
As the temporal evolution of the downward shift (to higher binding energy) on cooling and thermalizing essentially tracks the time-scale seen for a freshly-cleaved sample exposed to residual vacuum at low temperature, this $T$-dependent shift is most likely to be due to the reversible adsorption/desorption of residual gas molecules at the TI surface \cite{Jiang2012, King2011}.

\renewcommand\thefigure{A1}
\setcounter{figure}{0}

\begin{figure}[!b]
  \centering
  \includegraphics[width = 8.6 cm]{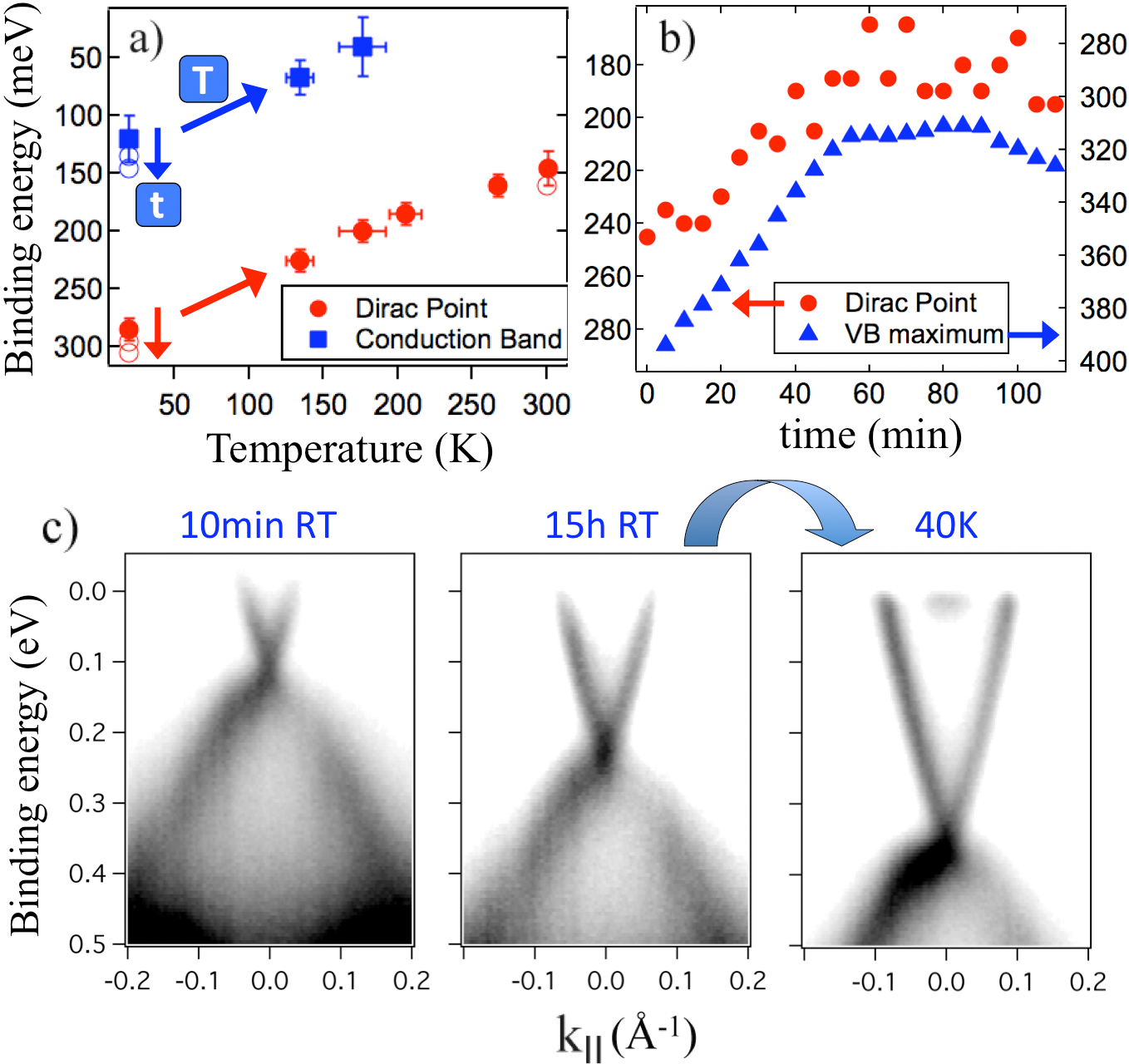}
  \caption{\textbf{The effect of temperature on the electronic band structure of Bi$_{1.5}$Sb$_{0.5}$Te$_{1.7}$Se$_{1.3}$ (BSTS1.5)} (a) $T$-dependent energy position of the Dirac point and the conduction band (CB) minimum on a temperature ramp from 20 to 300K (filled circles). After 18 hours at 300K, the data shown by the open green symbol were measured prior to re-cooling of the sample to 20K on which the binding energy of both the Dirac point and the CB minimum quickly recover to the previous, low-T values (uppermost open symbols at T=20 K). Both features continue shifting to higher binding energy as time passes (in this case a further 60 min), signalled by the small green arrows. (b) Time-dependent energy position of the Dirac point (left axis) and the valence band (VB) maximum (right axis) when the cryostat temperature is quickly increased to 150K but before full thermalization of the sample is attained. The error bars in the value of binding energy are estimated to be $\pm5$ meV for the valence band and range between $\pm10$ and $\pm15$ meV for the Dirac point. (c) Near-$E_{\textmd{F}}$ electronic band structure at 300K measured 10 min after cleavage (left), 15 hours after cleavage (middle) and upon subsequent cooling over 45 mins to 40K, followed by a thermalization period (waiting time) of 10mins (right panel).}
\label{figA1}
\end{figure}

Fig. A1a shows that for BSTS1.5, this means for temperatures above \textit{ca.} 200K, the CB in the near-surface region is above $E_{\textmd{F}}$ and thus depopulated.
The I($E,k$) image for BSTS1.5 shown in Fig. A1c illustrates that the surface band bending for this material at room temperature is insufficient to lead to occupation of the CB, even after 15h of residual gas exposure.
Nevertheless, for samples of macroscopic thickness the conductivity at higher temperatures will be dominated by impurity-related states \cite{Ren2010} which do not give rise to coherent, sharp bands in $k$-space.
Fig. A1c shows that on re-cooling to 40K, the ARPES signature changes significantly and only 10 minutes at low temperature are enough to cause the downward BB to increase by a further 100 meV, which once again brings the CB minimum below E$_{\textmd{F}}$, as seen in the right-hand panel of Fig. A1c.
Unlike the flattening of the band bending via the photo-induced effects described in the main body of the paper, the effects of changing the temperature are not local, but global in character.
This is the reason that experimental broadening of the I($E,k$) images like that seen in Fig. 3 is not observed in the data of Fig. A1c.

The data of Fig. A1 - ported to the world of transport experiments - mean that even in micron thin BSTS flakes in which the bulk contribution can be suppressed during a transport experiment \cite{Pan2014}, a second, topologically trivial conduction channel will open up in the SCR at low temperatures.
As a matter of fact, (1) adatom adsorption on a freshly cleaved sample held at low temperature and (2) cooling of a previously adatom-decorated sample will both lead to a fast downward energy shift with respect to the Fermi level, and to the occupation of the conduction band within the space charge region near the surface. 

The time-scale for this cooling-induced shift to higher binding energy (see Fig. A1a and the right two panels of Fig. A1c) is a lot faster than the time-scale of the return to downward BB seen after switching off the high-flux illumination that leads to band flattening (see the right-hand, grey-backed panel of Fig. 3a).
After cessation of high-flux illumination at low temperatures, the photo-induced upward energy shift decays very slowly and only partially, and this points to an important role for SPV in the band flattening due to the relatively long recombination lifetime of the electrons and holes created by the photon beam.
In contrast, on re-cooling, the bands shift back completely to higher binding energy, and do so at a rate resembling the fast, initial changes seen on the adsorption of residual gas atoms during the first few minutes of vacuum exposure at low $T$ on a freshly-cleaved sample (see, for example, the data of Fig. 2c).
These dynamics point to the mechanism of the fast, spatially global modification of the electronic structure observed on changing the sample temperature as being desorption/re-adsorption of residual gas atoms on increasing/decreasing the sample temperature.
This is in good agreement with results reported for both binary and ternary Bi-based TI materials \cite{Jiang2012, Zhu2011}.\\

\section*{APPENDIX 2: the effect of sample temperature on BSTS1.5 decorated with metal adatoms}

\renewcommand\thefigure{A2}
\setcounter{figure}{0}

\begin{figure} [!b]
  \centering
  \includegraphics[width = 8.1 cm]{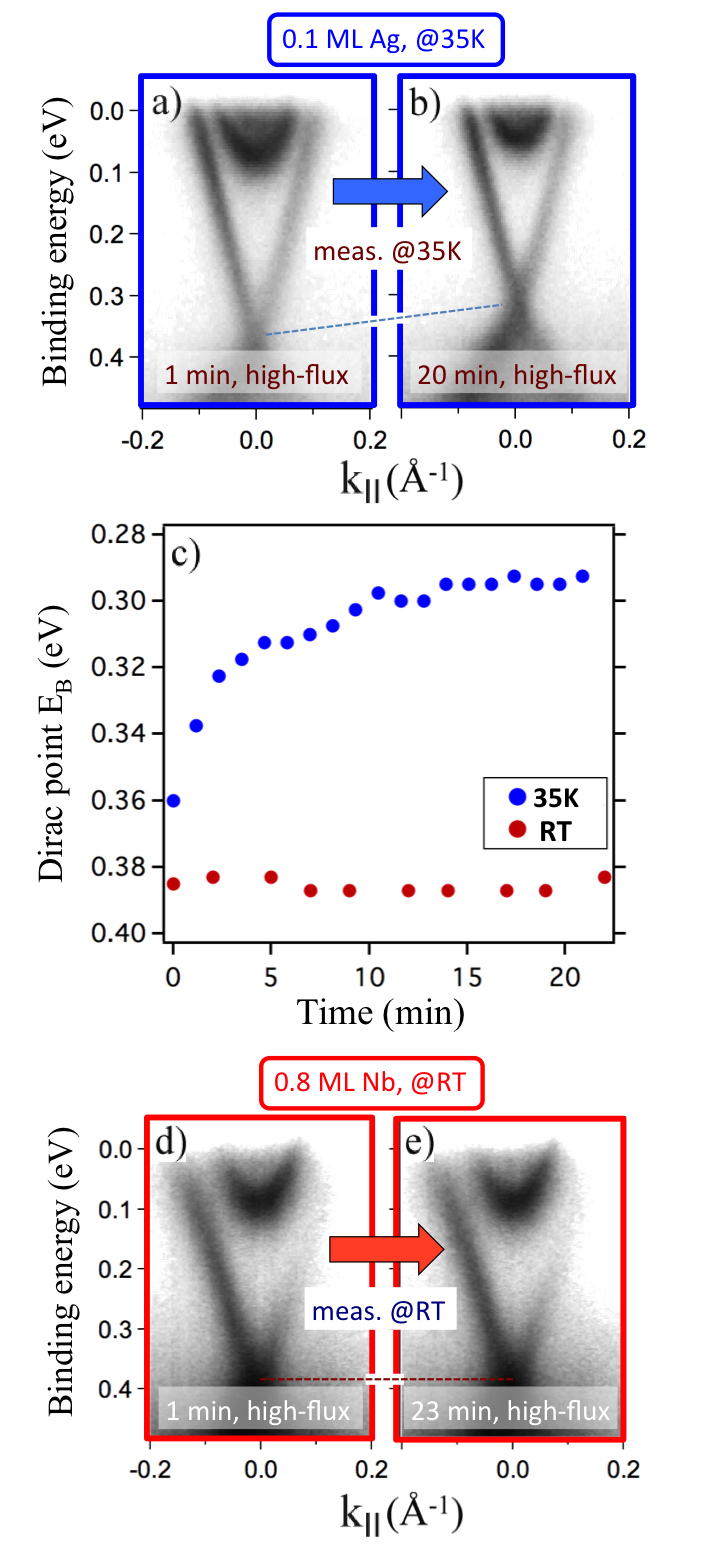}
  \caption{\textbf{
The effect of high-flux photons on Bi$_{1.5}$Sb$_{0.5}$Te$_{1.7}$Se$_{1.3}$ (BSTS1.5) band-bent due to metal adatoms and held at different temperatures.}
(a,b) The electronic band structure of BSTS1.5 after the deposition of 0.1 ML Ag at 35K. The spectra are acquired after (a) 1 min. and (b) 20 min. exposure to high-flux 27 eV photons at 35K, at which temperature the images were also measured.
(c) The evolution of the Dirac point binding energy for the same band-bent BSTS1.5 after exposure to high-flux photons at 35K (blue data points) and at RT (red data points). The error bars in the value of binding energy are estimated to be $\pm5$ meV. 
(d,e) The electronic band structure of BSTS1.5 after the deposition of 0.8 ML Nb at RT. The spectra are acquired at room temperature after (c) 1 min. and (d) 23 min. exposure to high-flux photons at RT.}
\label{figA2}
\end{figure} 

Here we present the effect of high-flux photons on different kinds of equivalently band-bent BSTS1.5 surfaces held at different temperatures.
As starting points, we achieved a comparable amount of downward band bending at room temperature (RT) and at 38K by evaporating 0.1 ML of Ag at 35K (blue data points) and 0.8 ML of Nb at RT (red data points).  
Details of the development of the electronic structure of the interfaces between metals and BSTS1.5 surfaces will be reported elsewhere.
Here it is sufficient to note that - since the total value of photon-induced energy shift depends on the amount of initial band bending - a comparable degree of downward band bending is important in order to perform a meaningful comparison between the situation at 35K and at room temperature.
The central panel of Fig. A2 shows that high-flux photons have practically no effect on the electronic band structure of band-bent BSTS1.5 held at RT (red data points).
This is different to the situation where the same photon fluence affects a similarly band-bent BSTS1.5 sample which is now kept at 35K (blue data points).
In the latter case, the energy of the Dirac point shifts to lower binding energies by around 70 meV after 20 minutes of photon exposure.
There are two important conclusions that can be made from these data.
Firstly, the qualitative effect of high-flux photons on BSTS1.5 does not depend on the origin of the initial band bending.
The same trends are seen, irrespective of whether the initial BB is a result of the adsorption of residual gas atoms from UHV (Figs. 1-4 in the main paper) or due to deliberate surface decoration with metals (Fig. A2).
Secondly, increasing the temperature to room temperature can effectively cancel the effect of high-flux photon exposure. 
This would be in keeping with strongly increased thermionic emission over the space charge region at high temperature resulting in an enhanced bulk-to-surface recombination current \cite{Kronik1999,Hecht1990,Chang1990}. 
The SPV mechanism provides a natural fit to both of these observations, since it depends on the sample temperature and the amount of initial band bending but not on whether it arises from residual gas adatoms or from intentional metal adatom evaporation onto the surface.

\end{document}